\documentclass{iopart}
\usepackage[pdftex]{graphicx}
\usepackage{iopams}
\begin{document}

\newcommand{\Evec}{\hbox{\boldmath{$\mathcal{E}$}}}
\newcommand{\Bvec}{\hbox{\boldmath{$\mathcal{B}$}}}
\newcommand{\Esca}{\mathcal{E}}
\newcommand{\Bsca}{\mathcal{B}}
\newcommand{\muvec}{\hbox{\boldmath{$\mu$}}}
\newcommand{\evec}{\vec{\epsilon}}

\title[Search for the electron electric dipole moment with ThO]{Search for the electric dipole moment of the electron with thorium monoxide}

\author{A C Vutha$^1$, W C Campbell$^3$, Y V Gurevich$^2$, N R Hutzler$^2$, M Parsons$^2$, D Patterson$^2$, E Petrik$^2$, B Spaun$^2$, J M Doyle$^2$, G Gabrielse$^2$ and D DeMille$^1$}

\address{$^1$ Department of Physics, Yale University, New Haven, CT 06520, USA}
\address{$^2$ Department of Physics, Harvard University, Cambridge, MA 02138, USA}
\address{$^3$ Joint Quantum Institute, University of Maryland and National Institute of Standards and Technology, College Park, MD 20742, USA}
\ead{amar.vutha@yale.edu}

\begin{abstract}
The electric dipole moment of the electron (eEDM) is a signature of CP-violating physics beyond the Standard Model. We describe an ongoing experiment to measure or set improved limits to the eEDM, using a cold beam of thorium monoxide (ThO) molecules. The metastable $H \ {}^3\Delta_1$ state in ThO has important advantages for such an experiment. We argue that the statistical uncertainty of an eEDM measurement could be improved by as much as 3 orders of magnitude compared to the current experimental limit, in a first-generation apparatus using a cold ThO beam. We describe our measurements of the $H$ state lifetime and the production of ThO molecules in a beam, which provide crucial data for the eEDM sensitivity estimate. ThO also has ideal properties for the rejection of a number of known systematic errors; these properties and their implications are described.
\end{abstract}


\section{Introduction}

Electric dipole moments of elementary particles have been the focus of experimental searches for over 50 years \cite{SPR57,FSB03}. The observation of a non-zero electric dipole moment of the electron (eEDM) would be evidence of CP-violation in the lepton sector, with deep implications for our understanding of particle physics and cosmology \cite{KL97,PR05}. The existence of an electric dipole moment directed along the spin of a particle requires non-invariance under both parity (P) and time-reversal (T). T-violation has been observed and measured in the decays of K- and B-mesons, and the results are all consistent with T-violation arising from a single source: the complex phase $\delta_{SM} \approx 0.99$ rad appearing in the quark mixing matrix in the Standard Model (SM) of particle physics \cite{YAA06}.  This phase gives rise to a non-zero predicted value of the eEDM $d_e$ in the SM: $|d_e| \lesssim 10^{-40} e$ cm \cite{KL97}, 13 orders of magnitude smaller than the current limit $|d_e| < 1.6 \times 10^{-27} e$ cm \cite{RCSD02}.  Despite the very small value of $d_e$ in the SM, it is widely expected that an observation of the eEDM is likely if the experimental sensitivity can be improved by a few orders of magnitude. The combination of new physics at the electroweak scale, plus order-unity T-violating phases, naturally gives rise to values of $d_e$ that are at or near the current experimental limit \cite{BCM+05}. Experimental limits on the size of $d_e$ already impose stringent constraints on Supersymmetry and other physics beyond the SM \cite{BCM+05,OPRS05}.

Experiments to search for the eEDM attempt to measure energy shifts arising from the interaction of an internal atomic or molecular electric field with a bound electron, a relativistic effect which can be enormously enhanced in some electronic states \cite{San65, MBD06, CJD07}. The current best limit on $d_e$ comes from the Berkeley experiments using atomic beams of thallium \cite{RCSD02}. Compared to atoms however, polar molecules offer the possibility of much larger internal electric fields that can be fully oriented with modest laboratory electric fields \cite{KL95}. 

For a single electron, the eEDM vector $\mathbf{d}_e = 2 d_e \mathbf{S}$, where $\mathbf{S}$ is the electron spin. (We set $\hbar = 1$ everywhere, for convenience of notation). In heavy molecules, where the relativistic effects that lead to large internal electric fields are important, spin-orbit effects tightly couple $\mathbf{S}$ with the electron's orbital angular momentum $\mathbf{L}$ to form the total angular momentum of the electron $\mathbf{J}_e = \mathbf{L} + \mathbf{S}$. Hence in such a system, the eEDM $\mathbf{d}_e$ lies along $\mathbf{J}_e$ on average. The total angular momentum of the molecule $\mathbf{J} = \mathbf{J}_e + \mathbf{R}$ is the sum of the electronic angular momentum $\mathbf{J}_e$ and the nuclear rotation $\mathbf{R}$. In the lab frame, the projection $\Omega \equiv \mathbf{J}_e \cdot \hat{n}$ of the electronic angular momentum onto the internuclear axis of the molecule $\hat{n}$ is a good quantum number. In a molecular state with a definite value of $\Omega$, the expectation value of the relativistic eEDM Hamiltonian $H_d$ \cite{MBD06} has the following form:
\begin{equation}
\langle H_{d} \rangle = -d_e\mathbf{J}_e \cdot \Evec_{mol} = - d_e \Esca_{mol} \ \Omega.
\end{equation}
The internal electric field $\Evec_{mol} = \hat{n} \Esca_{mol}$ is directed along the internuclear axis of the molecule. Note that the internal electric field $\Esca_{mol}$ as defined above is related to the effective electric field $\Esca_{\mathrm{eff}}$ that appears elsewhere in the literature (see \emph{e.g.} \cite{MBD06}) as follows: $\Esca_{\mathrm{eff}} = \Esca_{mol}\ \Omega$. 

In our experiment, molecules in a $J=1$ state are first polarized in a laboratory electric field $\Evec = \Esca\hat{z}$, which orients the internuclear axis $\hat{n}$ parallel to $\Evec$. The molecular angular momentum $\mathbf{J}$ is then prepared such that it is aligned perpendicular to $\hat{n}$. A small applied magnetic field $\Bvec \parallel \Evec$ induces a torque on the magnetic moment $\muvec =\mu \mathbf{J}$ of the state and causes the angular momentum vector $\mathbf{J}$ to precess at the Larmor frequency $\omega_p$. At the same time, the interaction of $\mathbf{d}_e$ with $\Evec_{mol}$ results in a shift of the precession frequency by an amount $\Delta \omega_p = 2 d_e \Esca_{mol} \Omega$. By changing the direction of $\Evec_{mol}$ relative to $\Bvec$, this shift will change sign if $d_e \neq 0$. The internal electric field $\Evec_{mol}$ can be reversed either by reversing the laboratory field $\Evec$ that polarizes the molecule, or by populating a molecular state with the opposite value of $\Omega$ (using the $\Omega$-doublet structure in the ThO molecule, as detailed below). After the molecule precesses in the electric and magnetic fields for a duration $\tau_c$, the direction of $\mathbf{J}$ is analyzed, yielding a signal proportional to the precession angle.

The shot noise-limited uncertainty in a measurement of an eEDM-induced shift $\Delta \omega_p$ of the precession frequency is given by 
\begin{equation}
\delta \omega_p = \frac{1/\tau_c}{\sqrt{N}}
\end{equation}
where $N$ is the number of molecules detected during the measurement. The statistical uncertainty in an eEDM measurement, $\delta d_e$, is therefore given by
\begin{equation}
\delta d_e = \frac{1/\tau_c}{2 \ \Esca_{mol} \sqrt{\dot{N} T}} \label{eq:stat_sensitivity}
\end{equation}
where $\dot{N}$ is the detection rate of molecules and $T$ is the integration time of the experiment. It is evident that the statistical sensitivity can be improved by increasing the coherence time $\tau_c$, the internal electric field $\Esca_{mol}$ in the molecular state, or the detection rate $\dot{N}$ of molecules in the experiment. We will point out how a molecular beam experiment with ThO can be used to gain in eEDM sensitivity by achieving high values for all of these quantities. 

This paper is structured as follows. After an overview of the measurement scheme in Section \ref{sec:measurement}, we describe how a cold beam of ThO molecules can be used to improve the statistical uncertainty of an eEDM measurement in Section \ref{sec:statistics}. Then in Section \ref{sec:systematics} we discuss the features of ThO that enable the rejection of a number of known systematic errors. Section \ref{sec:measurements} contains a description of some of our preliminary measurements on ThO. 

\section{Details of the measurement scheme} \label{sec:measurement}

\begin{figure}{
\centering
\includegraphics[width=0.62\columnwidth]{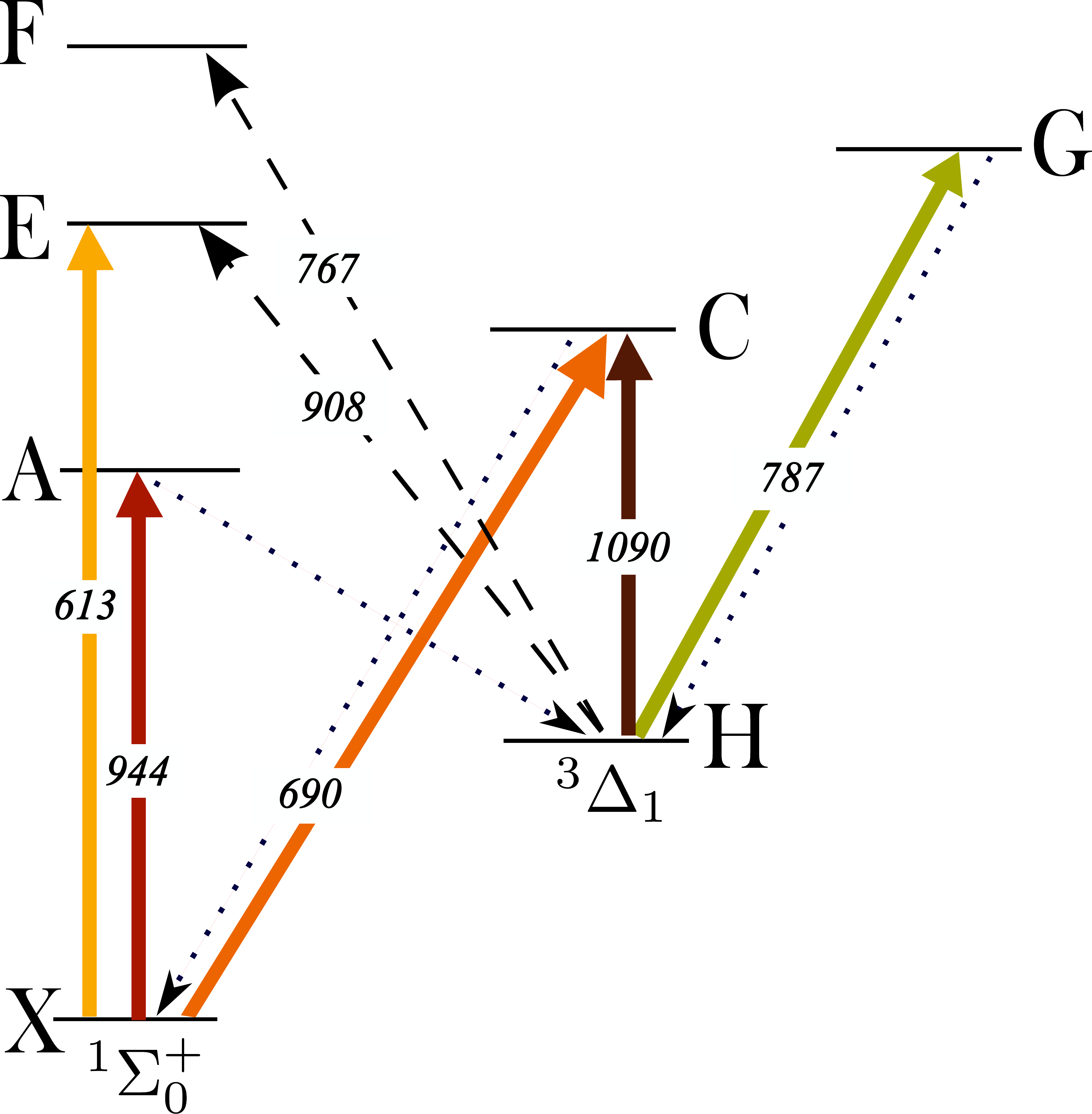}
\caption{Electronic states in the ThO molecule, shown schematically (based on \cite{EL85}). Bold arrows indicate transitions that we have observed using laser absorption or used to induce fluorescence. Dotted lines indicate transitions where spontaneous decay has been detected by laser-induced fluorescence or inferred from optical pumping effects. Dashed arrows indicate transitions that we hope to use in the future. Numbers on the arrows indicate their wavelengths in nm.} 
\label{fig:tho_levels}}
\end{figure}

\begin{figure}{	
\centering
\includegraphics[width=\columnwidth]{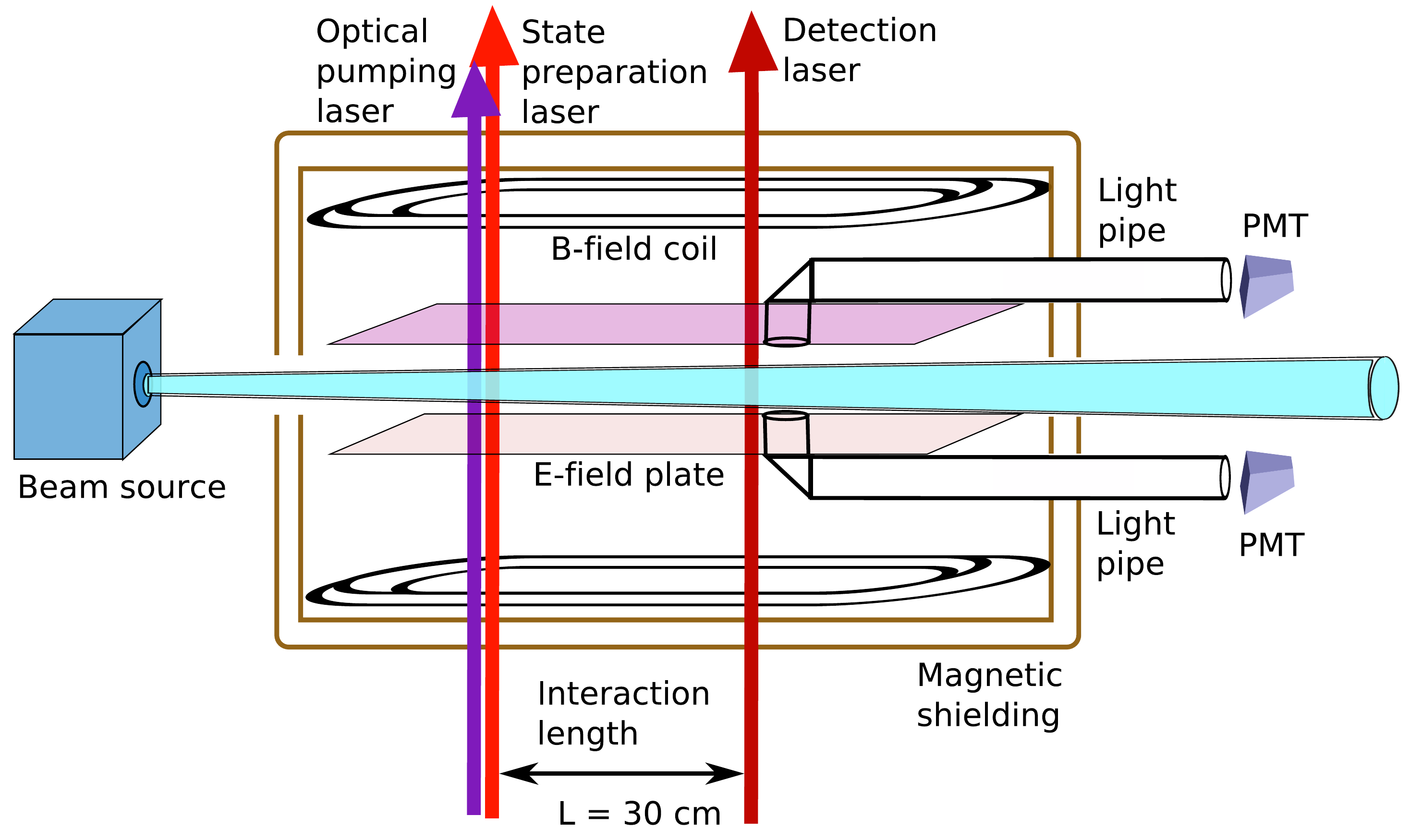}
\caption{Schematic of a molecular beam experiment to measure the eEDM using ThO. Cold molecules from the beam source enter a magnetically shielded interaction region where parallel electric and magnetic fields are applied. After freely precessing in the electric and magnetic fields over a 30 cm long flight path, the spin of the molecule is analyzed by coupling it to two orthogonal polarizations of the detection laser and detecting the resulting laser-induced fluorescence.} 
\label{fig:experiment}}
\end{figure}

\begin{figure}
	\centering{
	\includegraphics[width=0.62\columnwidth]{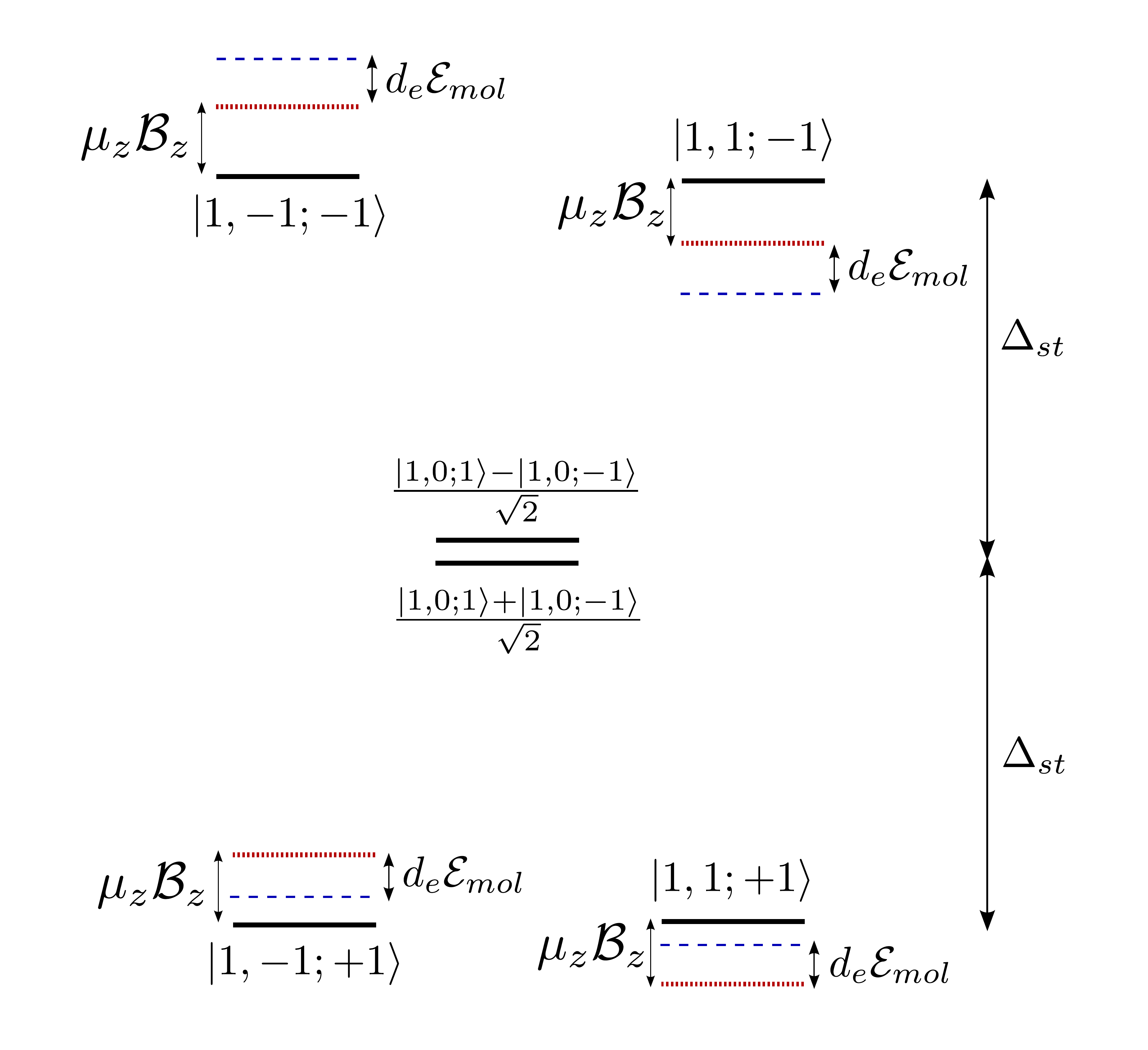}
	\caption{Schematic level structure and energy shifts of the ThO $H {}^3\Delta_1 (J=1)$ manifold of sublevels, in the presence of the applied external fields $\Evec = \Esca\hat{z}$ and $\Bvec = \Bsca\hat{z}$ (not to scale). States are labeled by their quantum numbers $|J,M_J;\mathcal{N}\rangle$, where $\mathcal{N} = \mathrm{sgn}\left( \hat{n} \cdot \Evec\right)$ describes the orientation of the fully polarized molecules. Note the different sign of the eEDM shift in the $\mathcal{N}=\pm 1$ $\Omega$-doublet levels.}
	\label{fig:omegadoublet}}
\end{figure}

Figure \ref{fig:tho_levels} shows some of the electronic states in the ThO molecule. The paramagnetic $H \ {}^3\Delta_1$ state is the locus of the eEDM measurement: it arises from a favourable combination of atomic orbitals in the constituent Th atom, and has a large value of the eEDM-enhancing internal electric field $\Esca_{mol}$. (Use of such ${}^3\Delta_1$ states for eEDM measurements was first suggested in \cite{SC04,MBD06}.) As shown schematically in Figure \ref{fig:experiment}, molecules from the beam source enter the interaction region and are first intercepted by an optical pumping laser tuned to the $X \to A$ transition. Excitation by this laser and subsequent $A \rightsquigarrow H$ spontaneous decay populates the $H$ state. All the states relevant to the measurement are in the ground rovibrational $v=0,J=1$ level of the $H$ state.  In the absence of an applied electric field $\Evec$, sublevels in this manifold are identified by their quantum numbers $M_J = \pm 1, 0$ (projection of $\mathbf{J}$ along $\hat{z} \parallel \Evec \parallel \Bvec$), and $P = \pm 1$ (parity). The opposite-parity $\Omega$-doublet levels in the $H$ state have a very small splitting ($\lesssim$ 10 kHz), which we neglect. With a sufficiently large applied field $\Evec$, the $P = \pm 1$ sublevels with the same value of $M_J$ mix completely; the resulting eigenstates have complete electrical polarization, described by the quantum number $\mathcal{N} \equiv \mathrm{sgn}\left( \hat{n} \cdot \Evec\right) = \pm 1$. (The $M_J=0$ sublevels do not mix.)  The relevant energy levels are shown in Figure \ref{fig:omegadoublet}. The tensor Stark shift $\Delta_{st}$ is defined as the magnitude of the shift of the oriented $|M_J| = 1$ levels from the unperturbed $M_J=0$ levels.

By coupling the molecules to the polarization of a strong state-preparation laser driving \emph{e.g.} the $H \to C$ transition, the molecule is initially prepared by depleting the coherent superposition of $|M_J=\pm1; \mathcal{N}\rangle$ that couples to the laser polarization $\evec_{sp}$, leaving behind a dark state. With the laser polarization $\evec_{sp} = \hat{y}$ for example, the initial state of the molecules becomes
\begin{equation}
|\psi_i^\mathcal{N}\rangle = \frac{|M_J = +1;\mathcal{N}\rangle + |M_J=-1;\mathcal{N}\rangle}{\sqrt{2}}
\end{equation}
with $\mathcal{N}=+1$ ($\mathcal{N}=-1$) corresponding to the lower (upper) $\Omega$-doublet component. The tensor Stark shift $\Delta_{st}$ is large enough that levels with different values of $\mathcal{N}$ are spectrally resolved by the laser. Hence a particular value of $\mathcal{N}$ is chosen by appropriate tuning of the laser frequency.

The molecules in the beam then travel through the interaction region, where the relative phase of the two states in the superposition is shifted by the interaction of the magnetic moment $\muvec_H$ with $\Bvec$, and $\mathbf{d}_e$ with $\Evec_{mol}$. After free evolution during flight over a distance $L$, the final wavefunction of the molecules is 
\begin{equation}
|\psi_f^\mathcal{N}\rangle = \frac{e^{i\phi} |M_J=+1;\mathcal{N}\rangle + |M_J=-1;\mathcal{N}\rangle}{\sqrt{2}}.
\end{equation}
For a molecule with velocity $v_f$ along the beam axis, the accumulated phase $\phi$ can be expressed as 
\begin{equation}
\phi = \int_{x=0}^{x=L} 2 \ (d_e \Esca_{mol} \Omega + \mu_H \Bsca) \ \frac{dx}{v_f} \equiv \phi_{\Esca} + \phi_B.
\end{equation} 

The phase $\phi$ will be detected by measuring populations in two ``quadrature components'' $|X_\mathcal{N}\rangle$ and $|Y_\mathcal{N}\rangle$ of the final state, where we define
\begin{eqnarray}
|X_\mathcal{N}\rangle & = & \frac{|M_J=+1;\mathcal{N}\rangle + |M_J=-1;\mathcal{N}\rangle}{\sqrt{2}}; \nonumber \\ 
|Y_\mathcal{N}\rangle & = & \frac{|M_J=+1;\mathcal{N}\rangle - |M_J=-1;\mathcal{N}\rangle}{\sqrt{2}}.  \label{eq:one}
\end{eqnarray}
The quadrature state $|X_\mathcal{N}\rangle$ ($|Y_\mathcal{N}\rangle$) can be independently detected, for example, by excitation with a laser whose polarization is $\evec_{d} = \hat{x}$ ($\evec_{d} = \hat{y}$).  The detected populations in the quadrature states, given by $P_{X} = |\langle X_\mathcal{N}|\psi_f^\mathcal{N}\rangle|^2$ and $P_{Y} = |\langle Y_\mathcal{N}|\psi_f^\mathcal{N}\rangle|^2$, can be expressed as $P_{X}  =  \frac{1 + \cos \phi}{2}$ and $P_{Y} = \frac{1 -  \cos \phi }{2}$.

The detected fluorescence signal from each quadrature state will be proportional to its population.  We express these signals ($S_X$ and $S_Y$) as a number of photoelectron counts per beam pulse, and write $S_{X(Y)} = S_0 P_{X(Y)}$, where $S_{0}$ is the total signal from one beam pulse.  To measure $d_e$, we adjust the magnetic field $\Bsca$ to yield a bias phase $|\phi_B| = \pi/2$ and construct the asymmetry $\mathcal{A}$:
\begin{eqnarray}
\mathcal{A} & = & \frac{S_Y-S_X}{S_Y+S_X} \cong  \mathrm{sgn}(\Bsca) \ \phi_{\Esca} \\
& \cong & 2 d_e \Esca_{mol} \ \frac{L}{v_f} \ \mathrm{sgn}(\Bsca) \cdot \mathrm{sgn}(\Esca) \cdot \mathrm{sgn}(\mathcal{N}). \label{eq:asym}
\end{eqnarray}
In the last step we have used the fact that our beam source has a narrow forward velocity distribution (with forward velocity $v_f$ and spread $\Delta v_\perp \ll v_f$), so that the offset phase $\phi_B$ is nearly constant for all molecules.  Note that $\mathcal{A}$ is odd in all three quantities $\Esca$, $\Bsca$, and $\mathcal{N}$, and is proportional to $\Esca_{mol}$.  The shot noise-limited uncertainty in $d_e$ for measurements on a single beam pulse, $\delta d_{e,0}$, can be derived from \eref{eq:stat_sensitivity} and \eref{eq:asym}:
\begin{equation} \label{eq:stat_sensitivity_2}
\delta d_{e,0} = \frac{1/\tau_c}{2 \ \Esca_{mol} \sqrt{S_0}}
\end{equation}
where the coherence time $\tau_c = L/v_f$.

The detection laser couples molecules in the $H$ state to an excited electronic state which decays to the ground electronic $X$ state with a large branching ratio (\emph{e.g.} the $E$ or $F$ state, see Figure \ref{fig:tho_levels}). This scheme allows for efficient rejection of scattered light from the detection laser, since the emitted fluorescence photons are at a much bluer wavelength than the laser. The spectroscopy of the ThO molecule has been extensively studied by previous workers \cite{MWGP87,EL84,GHKH05,WM97,HH79,PNH02}. As well as having spectroscopic constants that are precisely known for a number of electronic states, a significant experimental advantage of this system is provided by the fact that all the relevant transitions -- for optical pumping, state-preparation and detection -- are accessible with readily available laser diodes.

\section{Statistical sensitivity of a ThO beam experiment} \label{sec:statistics}

In this section, we describe the features of the ThO molecule that lead us to expect a significant increase in the statistical sensitivity to an eEDM. 

\subsection{Large internal electric field in ThO}

The $H \ {}^3\Delta_1$ state in ThO has a large enhancement of the effect of an eEDM. The application of a modest laboratory electric field $\Evec \leq$ 100 V/cm fully polarizes a ThO molecule in the $H$ state. Due to relativistic effects, this leads to a large internal electric field $\Esca_{mol}$  that is experienced by the valence electrons in the molecule \cite{MBD06,CJD07}. Recent calculations by Meyer and Bohn \cite{MB08} have verified that ThO has one of the largest values of the internal electric field for a diatomic molecule, $\Esca_{mol} \approx$ 100 GV/cm. The large internal electric field directly leads to enhanced sensitivity to an eEDM. 

\subsection{High flux beam source}

\begin{figure}
\centering
\includegraphics[width=0.62\columnwidth]{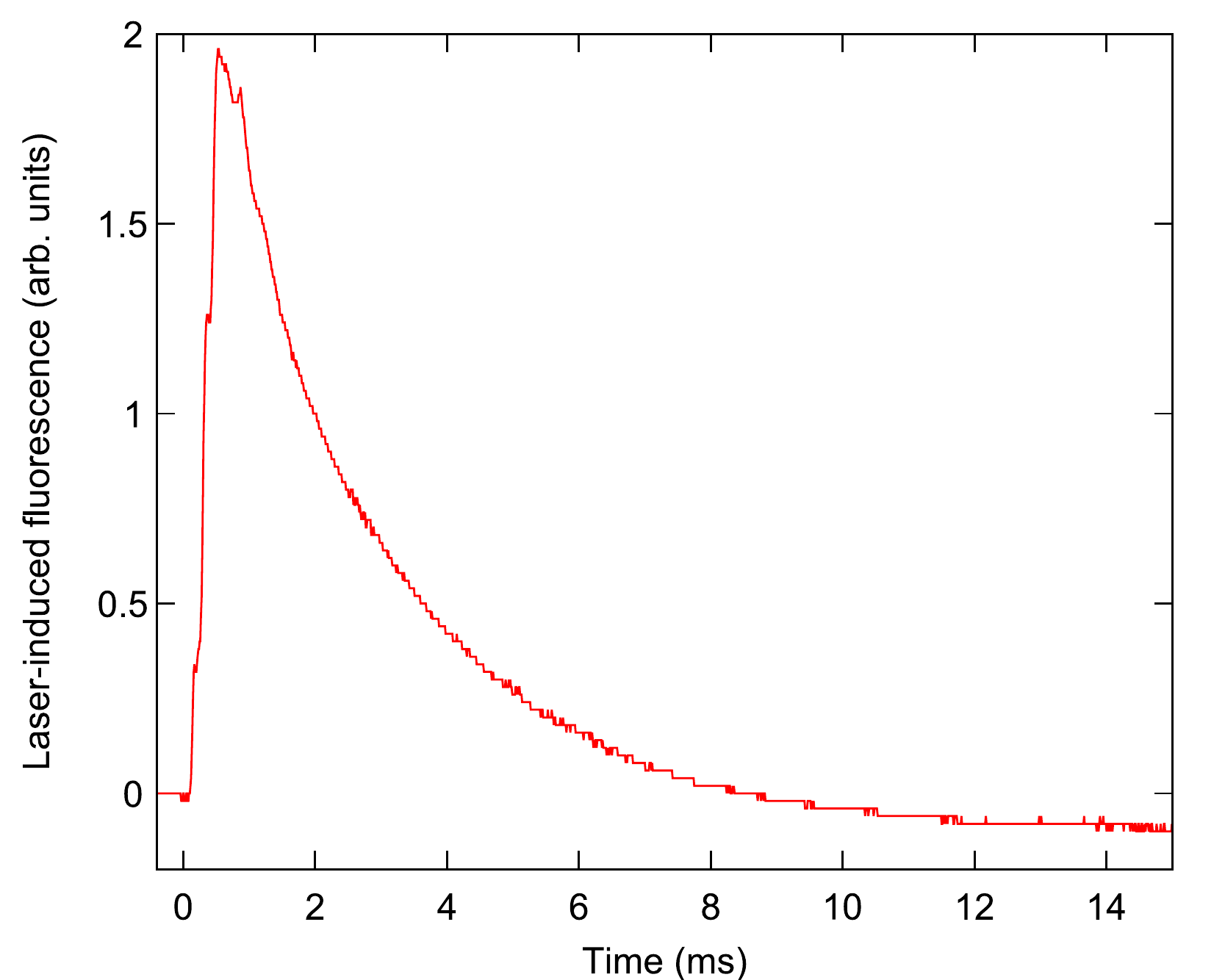}
\caption{Laser-induced fluorescence on the $X,v=0,J=1 \to C,v'=0,J'=1 \rightsquigarrow X$ transition at 14 489.98 cm$^{-1}$, observed in a molecular beam. Time $t=0$ corresponds to the firing of the ablation laser. The integrated fluorescence intensity, in combination with an estimate of the detection efficiency, yields the total number of molecules in this state per pulse of the beam.} 
\label{fig:beam}
\end{figure}

The detection rate of molecules in the experiment can be increased by using a beam source which provides a high flux of molecules in a single quantum state. Cryogenic buffer gas based sources operating in the intermediate hydrodynamic regime (with Reynolds number $\sim$ 10) have been shown to offer dramatic improvements in the intensity of seeded molecular species in a beam \cite{MBD+05,PD07,PRD09}. In our apparatus ThO molecules are produced by pulsed laser ablation of thorium dioxide (ThO$_2$) in a cryogenic cell held at 4 K. ThO molecules are cooled and hydrodynamically extracted into a beam by a continuous flow of helium buffer gas through the cell. The molecules in this type of beam have a small transverse velocity spread $\Delta v_\perp$ (characteristic of a $\sim$ 10 K Boltzmann distribution), but a larger forward velocity $v_f$ characteristic of a supersonic jet of helium, typically $v_f \approx$ 200 m/s. We expect that ThO beams made by this technique will have the same divergence as has been measured with other species such as Yb \cite{PD07}. We have made preliminary measurements to characterize our cold beams of ThO molecules. Figure \ref{fig:beam} shows a typical example of laser-induced fluorescence from the ground $X$ state of ThO on the $X \to C$ transition. Based on these measurements we estimate that $N_{beam} \approx 10^{10}$ molecules in a single quantum state are emitted into a solid angle $\Omega_b \approx 0.1$ sr during each beam pulse. With conservative estimates for the efficiency of optical pumping into the metastable $H$ state and the detection of laser-induced fluorescence photons, we estimate that the detection rate of molecules $S_0$ at the end of the interaction region could be $S_0 \sim 10^{5}$/s with a repetition rate of 10 Hz on the ablation laser. Details of the assumptions entering this estimate are given in Table \ref{tab:01}. 

\subsection{Long coherence time in the metastable $H$ state}

The radiative lifetime $\tau_H$ of the $H$ state sets the practical limit to the spin-precession coherence time $\tau_c$ in a molecular beam resonance experiment ($\tau_c \lesssim \tau_H$). The $H \ {}^3\Delta_1$ state in ThO is metastable, as the spin-flip transition to the $X \ {}^1\Sigma$ electronic ground state is nominally forbidden for an $E1$ transition. We have measured the lifetime of the $H$ state for molecules confined in a helium buffer gas cell (see Section \ref{sec:lifetime} for details of the measurement). Based on this measurement we estimate that the radiative lifetime $\tau_H \geq$ 1.8 ms. In practice, this is comparable to the time of flight $\tau_c$ of molecules through the interaction region, and we do not expect this to limit the statistical sensitivity of the experiment. 

\begin{figure}
\centering
\includegraphics[width=0.62\columnwidth]{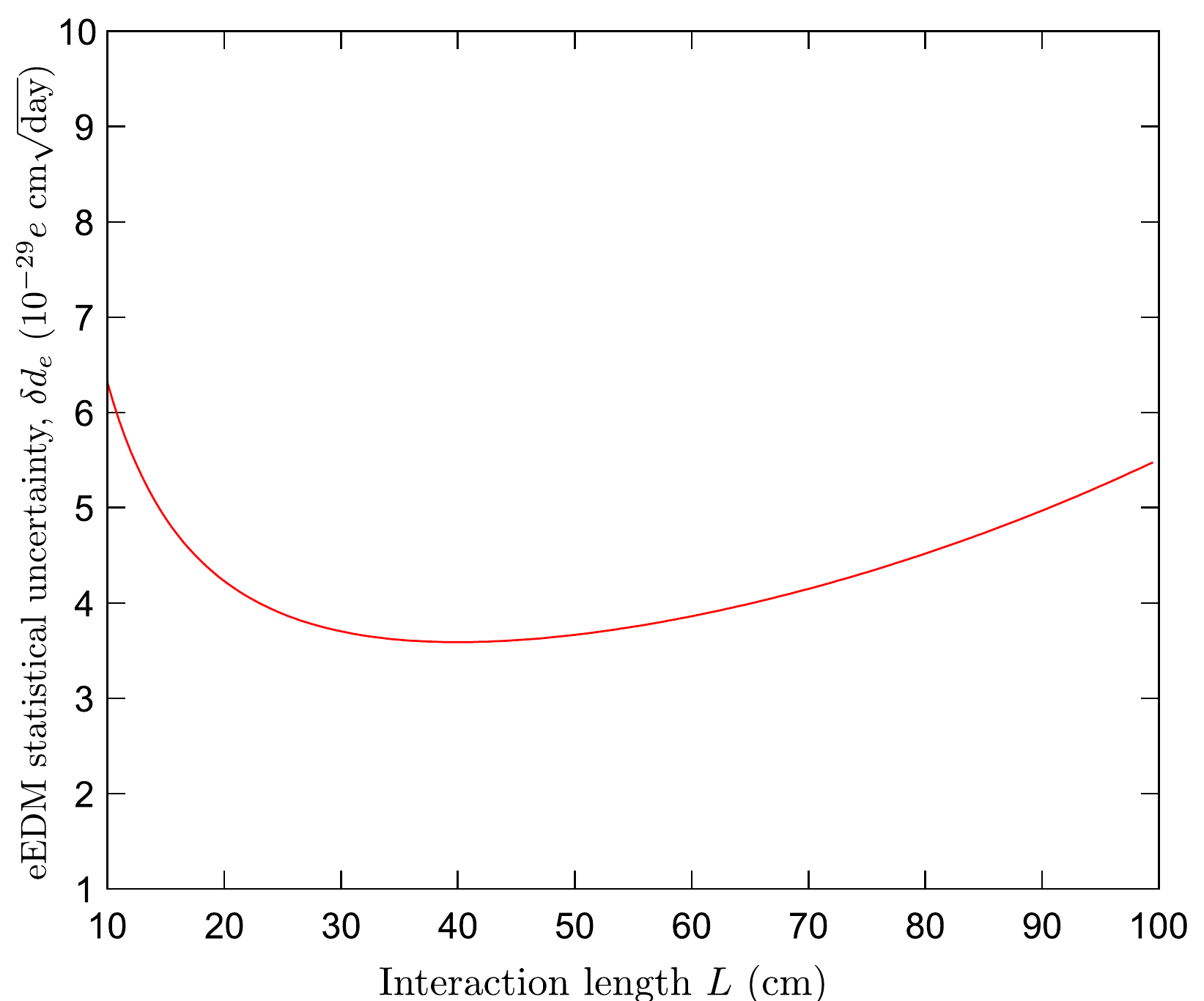}
\caption{Projected statistical uncertainty of the eEDM measurement as a function of the flight length in the interaction region $L$. Other parameters are the same as in Table \ref{tab:01}.} 
\label{fig:sensitivity}
\end{figure}

The resulting statistical uncertainty in the eEDM measurement is plotted in Figure \ref{fig:sensitivity}, where it is shown that our projection for the uncertainty of the eEDM measurement is quite insensitive to the time of flight in the apparatus. We emphasize that this is based on conservative estimates and known parameters, and only represents the promise of the simplest version of an experiment using ThO. We envision many future upgrades that could dramatically improve the statistical sensitivity in future generations of the experiment. 

\subsection{Technical noise sources}

Reaching the fundamental limit on the detection of the spin precession requires the suppression of technical noise to levels below the shot noise in the detection process. A number of technical noise sources of magnetic noise couple in via the Larmor precession of the magnetic moment of the molecule. The $H$ state in ThO, by virtue of its ${}^{2S+1}\Lambda_{\Omega} = {}^3\Delta_1$ configuration, is naturally insensitive to magnetic noise. The valence electrons in the $H$ state are in a spin triplet ($S = |\Sigma| = 1$) aligned anti-parallel to the orbital angular momentum projection ($|\Lambda| = 2$) to yield $|\Omega| = |\Lambda + \Sigma| = 1$. This configuration, combined with the fact that the spin $g$-factor is about twice as large as the $g$-factor for orbital angular momentum, leads to a net magnetic moment $\mu_H \equiv g_H \mu_B$ for the $H$ state that is nearly zero. (A nonzero value for $g_H$ arises due to spin-orbit mixing with other electronic states that have nonzero magnetic moments. We estimate $g_H \sim 0.01-0.10$ due to these effects.) This feature makes the experiment less susceptible to fluctuations in ambient magnetic fields. An unavoidable source of such fluctuations arises from thermal noise currents in the conductors surrounding the molecules (vacuum chamber, magnetic shields etc.). Based on analysis similar to \cite{Lam99,Mun05}, we estimate that the resultant frequency noise $\Delta \omega_{MJN}$ due to magnetic Johnson noise over the measurement bandwidth $1/\tau_c$ will be $\Delta \omega_{MJN} \leq 2\pi \times 30 \ \mu$Hz for our apparatus. In comparison, for the detection rate expected in the experiment, the shot noise-limited statistical uncertainty is $\Delta \omega_{SN} = 2\pi \times 80$ mHz. Other sources of technical noise in the precession frequency, such as from laser noise and fluctuations in the applied $\Bsca$ field, are also expected to be below shot noise levels.

It should be possible to detect spin precession with shot noise limited sensitivity, even in the presence of fluctuations in the molecular flux (due \emph{e.g.} to poor reproducibility of pulsed laser ablation). We intend to normalize the signal by measuring the asymmetry $\mathcal{A}$ (defined in \eref{eq:asym}) during every pulse of molecules. This will be accomplished by rapidly modulating the polarization of the detection laser $\evec_{d}$ between $\hat{x}$ and $\hat{y}$ directions and detecting the fluorescence from the molecules synchronously. By modulating the polarization of the detection laser much faster than the time scale over which the molecular flux varies within a single molecular beam pulse, we expect to reduce the noise due to shot-to-shot fluctuations in the molecular beam to levels below shot noise.

\begin{table}
  \begin{center}
    \begin{tabular}{ll}
	\hline
	Parameter & Estimate \\
	\hline ~ \\
	Beam yield: molecules/pulse/state	$N_{beam}$ & $10^{10}$ \\
	Beam forward velocity $v_f$ & 200 m/s \\
	Beam divergence $\Omega_b$		& 0.1 sr \\
	Beam length before interaction region $L_0$ & 50 cm \\
	State preparation efficiency	$e_p$		& 12 \% \\
	Flight length in interaction region $L$ & 30 cm \\
	Coherence time $\tau_c = L/v_f$ & 1.5 ms \\
	Surviving $H$ state fraction after flight & ~ \\
  \ \ \ $f = e^{-\tau_c/\tau_H}$ & 0.44 \\
	Detected area of beam $A$ & 1 cm$^2$ \\
	Detected solid angle of beam $\Omega_d$		& $1.5 \times 10^{-4}$ sr \\
	Geometric efficiency & ~ \\
	\ \ of fluorescence collection $e_g$ 	& 10 \% \\
	Quantum efficiency & ~ \\
	\ \ of photon detection	$e_q$	& 10 \% \\
  Photon counts/pulse & ~ \\
   \ \ \ $S_0 = N_{beam} f e_p e_g e_q \frac{\Omega_d}{\Omega_b}$ & $8.2 \times 10^{3}$ \\
	Internal field in ThO (from \cite{MB08}) $\Esca_{mol}$ & $2.5 \times 10^{25}~\mathrm{Hz}/e \  \mathrm{cm}$ \\
	eEDM uncertainty (single pulse) $\delta d_{e,0}$ & $2.4 \times 10^{-26} e \ \mathrm{cm}$\\
	Ablation laser repetition rate	$R$	& 10 Hz \\
	\hline
	eEDM uncertainty	$\delta d_e$	& $3.7 \times 10^{-29} e \ \mathrm{cm}/\sqrt{D}$\\
	($D$ days @ 50\% duty cycle) & ~ \\
	\hline
	\end{tabular}
\caption{Estimated eEDM statistical sensitivity of a first-generation ThO experiment, using   \eref{eq:stat_sensitivity_2} and our measured values/best estimates of various parameters.} \label{tab:01}  
  \end{center}
\end{table}

\section{Systematic error rejection with ThO} \label{sec:systematics}

In this Section, we describe our analysis of known systematic errors that could affect the measurement. We find that the choice of the $H \ {}^3\Delta_1$ state in ThO enables the rejection of known systematic errors to a level that is better than the projected statistical sensitivity of the measurement. $\Omega$-doubling in the $H$ state is an important feature of ThO that allows this rejection. $\Omega$-doublets are closely spaced levels of opposite parity that arise from the interaction between the electronic orbital angular momentum and the rotation of the nuclei in the molecule \cite{BC03,DBB+00}. These levels are easily mixed by a laboratory electric field and lead to a very large polarizability (and consequently, large tensor Stark shifts). In addition, the internal electric field $\Evec_{mol}$ has equal magnitude and opposite sign in the different components of the doublet \cite{DBB+00,KBB+04}, allowing us to reverse the internal electric field spectroscopically, without reversing the laboratory electric field. As explained in the following section, comparison of the spin precession signal between the two $\Omega$-doublet components allows the rejection of a number of known systematic errors. We find that all known sources of systematic effects, with conservative estimates of experimental imperfections, give rise to systematic errors $\delta d_e (\textrm{sys}) \ll 10^{-32} e \ \textrm{cm}$. 

\subsection{Magnetic effects}

Many typical systematics that affect molecular beam eEDM experiments involve spurious magnetic fields, arising \emph{e.g.} from leakage currents across the electric field plates, or from magnetic fields in the molecules' rest frame due to their motion in the static electric field. We do not expect either of these effects to be significant for the ThO experiment. Magnetic fields due to leakage currents are naturally small here since the electric field required to fully polarize the ThO molecule is $\lesssim$ 100 V/cm. Moreover, as discussed above the $H \ {}^3\Delta_1$ state has a suppressed magnetic moment. Finally, since the pair of $\Omega$-doublet states have nearly identical magnetic moments \cite{BHJD08}, any effects due to leakage currents largely cancel in the comparison between these levels. With a conservative estimate of leakage current ($I_{leak} \sim 1$ pA), we estimate systematic errors due to leakage currents to satisfy $\delta d_e (\textrm{sys}) \ll 10^{-32} e \ \textrm{cm}$.

An important advantage of polar molecules for eEDM searches is due to the large tensor Stark splitting in an electric field (expected to be $\sim 30-100$ MHz for our experimental parameters). Since the motional magnetic field ($\sim 10^{-10}$ T) is transverse to the laboratory electric field and because of the symmetry of tensor Stark splitting, its effect only appears as a third order perturbation and is further suppressed by the large tensor Stark energy \cite{PS70}. We estimate systematic errors due to motional fields to satisfy $\delta d_e (\textrm{sys}) \ll 10^{-32} e \ \textrm{cm}$. 

\subsection{Electric quadrupole shifts}

The measurement takes place in the lowest rovibrational level of the electronic $H$ state in ThO, in the $\Omega$-doubled level with total angular momentum $J = 1$. The nonzero electric quadrupole moment in the $J=1$ level can couple to electric field gradients and lead to a frequency shift between the $|M_J = +1\rangle$ and $|M_J = -1\rangle$ Zeeman sublevels that can mimic an eEDM under some circumstances. The Hamiltonian for the electric quadrupole interaction is $H_Q = \sum_{ij} Q_{ij} \nabla_i \Esca_j$, 
where $Q_{ij}$ is the quadrupole moment of the state. We base our estimates on a typical size of $Q_{ij} \sim 1~ ea_0^2 \sim 10 ~$mHz/(V/cm$^2$). The most dangerous systematic effect would arise from an electric field gradient $\partial \Esca_x / \partial y$, which due to $H_Q$ gives rise to an off-diagonal interaction coupling $|M_J = +1\rangle$ with $|M_J = -1\rangle$.  The energy shift due to this off-diagonal quadrupole interaction adds in quadrature with the Zeeman splitting between these levels. The precession frequency thus acquires an additional shift
$\Delta \omega_Q = \ \frac{H_Q^2}{2 \mu_H \Bsca}$.  
$H_Q$ can generate a systematic error that survives apparent reversal of $\Evec$, if there are simultaneously field gradients $\partial \Esca^r_x / \partial y$ that reverse when the applied voltage is changed (e.g.\ due to tilts in the field plates) and gradients $\partial \Esca^{nr}_x / \partial y$ that do not reverse with the applied voltage (e.g.\ from patch potentials on the plates).  We estimate such effects using $\partial \Esca^r_x / \partial y$ = 10 mV/cm$^2$ and $\partial \Esca^{nr}_x / \partial y$ = 10 mV/cm$^2$. Even before taking into account the cancellation of this effect in the comparison between $\Omega$-doublet components, electric quadrupole effects generate a negligible systematic effect $\delta d_e (\textrm{sys}) \ll 10^{-32} e \ \textrm{cm}$.

\subsection{Geometric phases}

Spatial inhomogeneities in the applied electric and magnetic fields, which appear as time-varying fields in the rest frame of molecules in the beam, can give rise to geometric phase-induced systematic effects \cite{Ram55,Com91}. We have used the formalism outlined in \cite{VD09} to analyze geometric phase effects in an $\Omega$-doublet state in a polar molecule, in a combination of time-varying electric and magnetic fields. We briefly describe the details of the calculation here. The structure of a $J=1$ $\Omega$-doublet in a combination of static electric and magnetic fields is shown in Figure \ref{fig:omegadoublet}. Similar to the analysis in \cite{VD09}, we decompose the electromagnetic fields experienced by the molecules into $\hat{z}$-directed static components $\Esca_z, \Bsca_z$ and time-varying transverse components $\Esca_\perp, \Bsca_\perp$ in the $xy$-plane. We analyze the effect for single Fourier components of the evolving transverse fields, where the transverse electric (magnetic) field rotates at a frequency $\omega^E_\perp$ ($\omega^B_\perp$).

The transverse fields $\Esca_\perp$ ($\Bsca_\perp$) couple $|J,M_J=\pm 1\rangle$ to $|J,M_J=0\rangle$ and lead to AC Stark (Zeeman) shifts that result in geometric phases. For simplicity, we only analyze the interaction of the $|M_J = \pm 1\rangle$ in the two different $\Omega$-doublet components with one of the unshifted $|M_J = 0\rangle$ levels. Using an analysis similar to that in \cite{VD09}, we can write the off-resonant energy level shifts as follows:
\begin{eqnarray}\label{eq:omegadoublet}
\Delta E_{\pm 1}^{(u)} = \Delta E_{\pm 1}^{(l)} & = & \frac{1}{2} \Big[\frac{\mu_\perp^2 \Bsca_\perp^2}{\Delta_{st} + \mu_z\Bsca_z - \omega^B_\perp} + \frac{d_\perp^2 \Esca_\perp^2}{\Delta_{st} + \mu_z\Bsca_z - \omega^E_\perp} \nonumber \\
& & - \frac{\mu_\perp^2 \Bsca_\perp^2}{\Delta_{st} - \mu_z\Bsca_z + \omega^B_\perp} - \frac{d_\perp^2 \Esca_\perp^2}{\Delta_{st} - \mu_z\Bsca_z + \omega^E_\perp} \Big].
\end{eqnarray}

Here $\Delta E_{\pm 1}^{(u)}$ ($\Delta E_{\pm 1}^{(l)}$) is the additional energy shift between the $|M_J=\pm1\rangle$ states in the upper (lower) component of the $\Omega$-doublet due to the off-resonant electromagnetic fields. We have defined the transverse matrix elements $\mu_\perp$ ($d_\perp$) of the magnetic (electric) dipole moment between the $|M_J = \pm1\rangle$ states and the $|M_J=0\rangle$ state.  \eref{eq:omegadoublet} is valid up to second order in the transverse fields $\Esca_\perp, \Bsca_\perp$. We find that the off-resonant energy shift which leads to geometric phases is identical for both the components of the $\Omega$-doublet. The fact that geometric phase shifts are identical for the two $\Omega$-doublet components does not rely on the assumption of adiabatic evolution of the transverse fields in the rest frame of the molecules, making this result quite general. 

Due to the large tensor Stark splitting in polar molecules such as ThO, magnetic effects are suppressed and the geometric phase is dominated by electric field effects. Uncontrolled transverse components $\Esca_\perp$ of the electric field, through a combination of non-reversing components (such as from patch potentials, $\Evec^{nr} \sim$ 10 mV/cm) and reversing components (\emph{e.g.} from misalignments of the field plates, $\Evec^r \sim$ 10 mV/cm) can give rise to geometric phase shifts for a single $\Omega$-doublet component. With these typical values, the geometric phase could lead to a systematic error $\delta d_e (\textrm{sys}) \sim 4 \times 10^{-31} e \ \textrm{cm}$.  Crucially however, the geometric phase picked up between $|M_J = \pm 1\rangle$ sublevels is the same for both the components of the $\Omega$-doublet. Therefore, comparison between the $\Omega$-doublet components in the $H$ state of ThO will allow us to reject systematic errors due to geometric phases, and we expect that the residual effect of geometric phases should give $\delta d_e (\textrm{sys}) \ll 10^{-32} e \ \textrm{cm}$. As an aside, we point out that the rejection of geometric phase systematic errors using $\Omega$-doublets could be useful for eEDM experiments with trapped polar molecules \cite{SBL+07,RS08,THSH09}. 

\section{Preliminary measurements} \label{sec:measurements}

\subsection{ThO molecules in a buffer gas cooled beam}

We have measured the flux of cold molecules in a buffer gas cooled molecular beam of ThO, using a  simple test beam source. ThO molecules are produced by pulsed laser ablation of solid ThO$_2$ using $\sim$ 5 ns long, 10 mJ pulses at 532 nm from an Nd:YAG laser. The ablation target is a sintered pellet of ThO$_2$ powder with a density of 6.9 g/cm$^3$ \footnote{This target was fabricated at Oak Ridge National Laboratory, and kindly provided to us by Dr. Daniel Stracener.}. The pellet is held inside a copper cell which is thermally anchored to a liquid helium cryostat. Helium gas continuously flows at a rate of $4 \times 10^{18}$ atoms/s through the cell and exits it through a 1 mm $\times$ 6 mm aperture in a 0.25 mm thick copper plate. We probe molecules outside the cell in a region where the vacuum is maintained at $\sim 10^{-6}$ Torr by cryopumping. Laser-induced fluorescence is collected after exciting molecules from low-lying rovibrational states of the ground electronic $X \ {}^1\Sigma$ state to the $C$ state (see Figure \ref{fig:tho_levels}). Figure \ref{fig:beam} shows a typical fluorescence trace. In this test apparatus, we typically observe a pulse of $N_{beam} \approx 10^{10}$ molecules in a single quantum state spread out over $\sim$ 5 ms. The number of molecules is determined from the signal size by estimating the collection efficiency of the fluorescence signal; this number is also separately verified by direct measurements of laser absorption on the $X \to C$ transition. This source has been operated at repetition rates up to 15 Hz. We are in the process of constructing a full-scale beam apparatus, based around a cryogen-free refrigerator, that is capable of higher cooling power and a longer duty cycle. We expect that this new apparatus could provide a larger flux of cold molecules with a higher repetition rate and/or total integrated flux. We have also fabricated dense ThO$_2$ ablation targets using the Nb$_2$O$_5$ doping method described in \cite{BVK+88}, and produced void-free, homogeneous ceramics. We are working to optimize them for  ablation yield and target longevity.

\begin{figure}
\centering
\includegraphics[width=0.62\columnwidth]{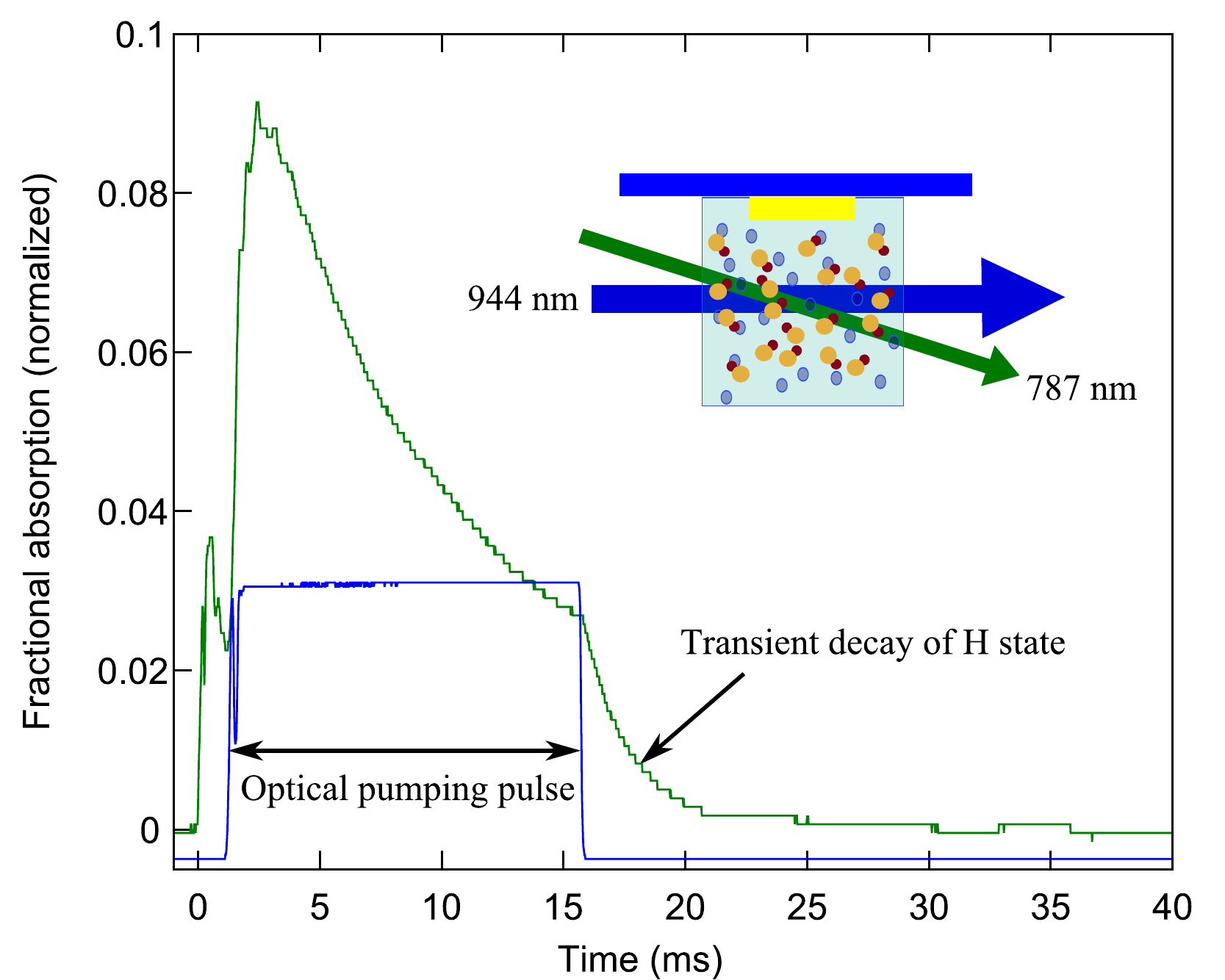}
\caption{Measurement of the $H$ state lifetime. For this data the $H$ state population was monitored by laser absorption on the $H \to G$ transition. The probe laser was overlapped inside the buffer gas cell with an optical pumping beam exciting the $X \to A \rightsquigarrow H$ transition, as shown schematically in the inset. An exponential fit to the transient decay of the absorption signal yields a lifetime $\tau_{cell}$ = 1.8 ms.} 
\label{fig:lifetime}
\end{figure}

\subsection{Measurement of the lifetime of the $H$ state} \label{sec:lifetime}

We have measured the lifetime of ThO in the metastable $H$ state using a closed cryogenic buffer gas cell (see Figure \ref{fig:lifetime}). The apparatus is similar to that described in the previous section, with the exception that the cell used for this measurement had no aperture and no flow of helium gas. The cell was filled with about 50 mTorr of helium gas at $\simeq$ 4 K and laser ablation in the buffer gas environment produced $\sim 10^{11}$ ThO molecules in the ground $X$ state. Molecules from the $v=0,J=1$ level in the ground $X$ state were optically pumped into the $H$ state via the $A$ state (see Figure \ref{fig:tho_levels}). This was accomplished by illuminating them with 10 mW of light, resonant with the $X, v=0, J=1 \to A, v'=0, J'=0$ transition at 10 600.15 cm$^{-1}$, for a variable duration. Molecules in the $H$ state were probed by laser absorption on the $H,v=0,J=1 \to G,v''=0,J''=2$ transition at 12 694.58 cm$^{-1}$. We observed that the $H$ state was continuously repopulated, at a low level, following the ablation pulse. (This is believed to be due to radiative or collision-induced decays from higher-lying electronic states and/or vibrational levels populated during the ablation pulse.) To eliminate errors due to this repopulation, we only analyzed the transient decay of optically pumped molecules in the $H$ state after the optical pumping beam was rapidly extinguished (in $<$ 500 $\mu$s). Exponential fits to the time decay yielded lifetimes which were independent of the buffer gas density in the cell within our experimental resolution. We expect that the observed decay lifetime $\tau_{cell}$ = 1.8 ms is a lower limit to the radiative lifetime $\tau_H$ of the $H$ state. We note as well that in these experiments, the $X$ state population was monitored via laser absorption on the $X \to C$ transition. We observed strong depletion of the $X$ state population when the $X \to A$ laser was on. Since the $A$ state can only decay to the $H$ or $X$ states, this indicates a high efficiency for optical pumping from $X$ into $H$. We are now working to quantify this efficiency. 

\subsection{Observation of blue-shifted fluorescence from the $H$ state}

\begin{figure}
\centering
\includegraphics[width=0.62\columnwidth]{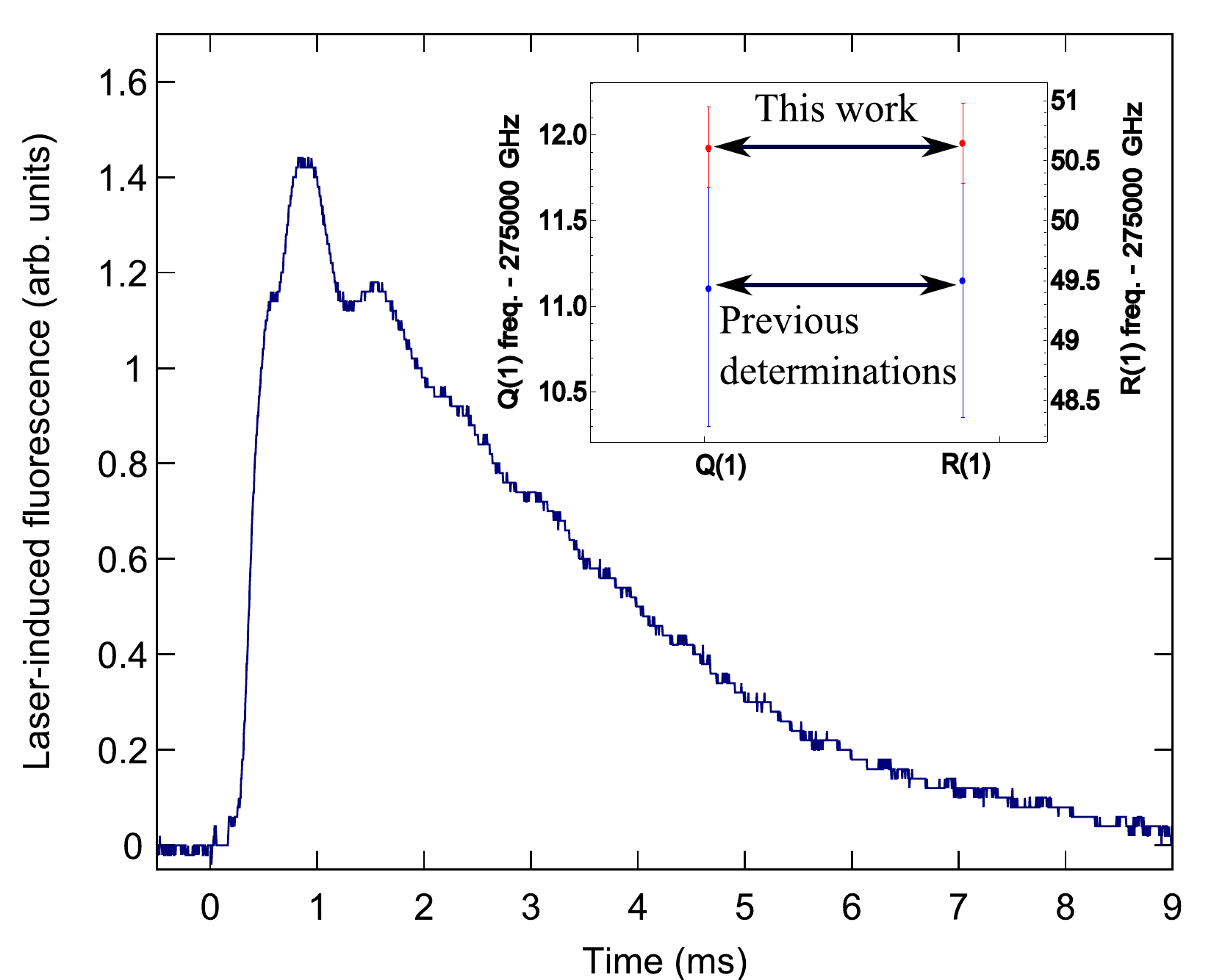}
\caption{Laser-induced fluorescence on the $H,v=0,J=1 \to C,v'=0,J'=1 \rightsquigarrow X$ transition at 9 173.38 cm$^{-1}$, after first optically pumping molecules from $X \to A \rightsquigarrow H$. Time $t=0$ corresponds to the firing of the ablation laser. The inset shows our measurement of the transition frequencies of the $Q(1)$ and $R(1)$ lines on the $H \to C$ transition, which had not been directly observed before. These are in good agreement with, and more accurate than, the splitting inferred from other measurements of the absolute $H$ and $C$ state spectroscopic constants \cite{EL85}.} 
\label{fig:blueshift}
\end{figure}

As a first step towards the detection of molecules in the $H$ state using blue-shifted fluorescence, we have excited ThO molecules on the $H \to C$ transition at $\sim$ 9 200 cm$^{-1}$ in a molecular beam. The $H$ state was populated by optically pumping molecules from the ground state, via laser excitation of the $X \to A$ transition and subsequent $A \rightsquigarrow H$ decay. A large fraction of these molecules excited to the $C$ state emit photons at 690 nm and decay to the $X$ state. Figure \ref{fig:blueshift} shows a typical time trace of the laser-induced fluorescence collected on this transition. We have observed saturation of this transition with a laser intensity $I_{H-C} \approx 10$ mW/cm$^2$, indicating that the $H$ state can be probed via this transition with high probability to emit one photon per molecule. In the future, we plan to explore the possibility of probing the $H$ state with other transitions such as $H \to E$ or $H \to F$, where the emitted fluorescence will be at wavelengths more suitable for efficient detection using photomultiplier tubes. 

\section{Summary}

We have described the launch of an experiment to search for the electric dipole moment of the electron, using cold ThO molecules in a buffer gas cooled beam. The $H$ state of ThO is a very promising system for achieving unprecedented statistical sensitivity in such a measurement, in addition to having ideal features for the rejection of known systematic errors. The combination of novel beam source technology and the experimental advantages offered by ThO could potentially lead to a significantly improved measurement of the eEDM. We have argued that in a first-generation experiment using ThO, integration for $D$ days could yield a statistical sensitivity $\delta d_e \sim 4 \times 10^{-29}/\sqrt{D} \ e \ \textrm{cm}$. A variety of future upgrades (including more efficient state-preparation and detection schemes, improved source flux, molecular beam focusing etc.) could  yield further improvements. 

\section*{Acknowledgments}
We acknowledge the technical assistance of Jim MacArthur and Stan Cotreau in the construction of apparatus. A.V. and D.D. acknowledge stimulating discussions with Paul Hamilton during the early stages of the project. D.D. thanks Dmitry Budker for discussions about systematic effects in eEDM measurements using $J=1$ systems. This work was supported by the US National Science Foundation.

\section*{References}
\bibliography{tho}

\end{document}